# Eliminating zeroth-order light of spatial light modulator with voltage optimization

Yueqiang Zhu, Kaige Wang, Jintao Bai, Wei Zhao*

[1]State Key Laboratory of Photon-Technology in Western China Energy, International Collaborative Center on Photoelectric Technology and Nano Functional Materials, Institute of Photonics & Photon Technology, Northwest University, Xi'an 710127, China

* Correspondence: zwbayern@nwu.edu.cn

The crucial zeroth-order light due to the pixelation effect of spatial light modulator (SLM) has been a serious issue in the field of light modulation, especially in applications with a high numerical aperture optical system. In this investigation, we report that by properly adjusting the high-level and low-level pixel voltages of an SLM, the zeroth-order light caused by the pixelation effect of SLM can be significantly eliminated. The method is further validated in an inverted fluorescence microscope. The experimental results show that the zeroth-order light can be inhibited up to 91.3%, accompanied by an improvement of the modulation efficiency from 77.5% to 92.6%.

Nowadays, spatial light modulator (SLM) has attracted long-term interest in the development of both science and techniques[1-4], to reach flexible beam shaping with desired polarization, intensity, and phase distribution. It can generate diverse light distributions, from simple and regular light spots to complex and flexible 3D light fields with arbitrary intensity and polarization distributions [5-8]. For instance, SLM is used in the field of laser fabrication to process bionic structures and then simulate the living environment of yeast cells [9]. Besides, SLM is also widely used in laser fabrication [3], optical communication [4, 10], optical storage [11, 12], optical trapping [13, 14], optical imaging [15] and etc.

However, even if the computer-generated hologram (CGH) is well-designed, the zeroth-order light according to the pixelation effect of the SLM liquid crystal array is inevitable. Since the power density of the zeroth-order light can be much higher than that of the modulated beams [16, 17], this becomes especially terrible in the field of laser fabrication with SLM, where the stubborn zeroth-order light can seriously affect the fabricated structures. Therefore, the elimination of the zeroth-order light of SLM is of great interest.

To eliminate the influence of zeroth-order light, many efforts have been made, e.g., applying blazed grating. This can make the modulated beams deviate from zeroth-order light, which is subsequently shielded by an iris diaphragm [18] or other optical elements [19]. The zeroth-order light can also be moved away from the modulated beam along axial direction through a spherical lens phase [20], or inhibited with destructive interference [21]. However, these methods are less effective in a high numerical aperture (NA) optical system.

In this investigation, we propose a method that, by properly adjusting the high-level ($V_H$) and low-level ($V_L$) pixel voltages of a phase-only liquid crystal SLM, the zeroth-order light caused by the pixelation effect of SLM can be significantly eliminated, without additoinal shielding. More attractive is, accompanied by the elimination of zeroth-order light, the modulation efficiency can be improved simultaneously.

The phase modulation of a phase-only SLM is realized through controlling the tilting angle of liquid crystal (LC) cells, which is determined by the applied voltage. For a twisted nematic liquid crystal cell, initially there is no LC cell tilting if no voltage applied. The tilting angle of liquid crystal molecule $\varphi = 0$. When a voltage $V$ is applied, $\varphi$ can be expressed as [22]

$$\varphi = \begin{cases} 0, & V \leq V_c \\ \frac{\pi}{2} - 2\tan^{-1}\left[\exp\left(-\frac{V-V_c}{V_0}\right)\right], & V > V_c \end{cases} \quad (1)$$

where $V$ is the applied voltage on the LC, $V_c$ and $V_0$ are the threshold voltage and the excess voltage, respectively.

The relationship between $\theta = \pi/2 - \varphi$ ($\theta$ is the angle between the propagation direction of incident light and the liquid crystal molecule) and the equivalent extraordinary refraction index $n_e(\theta)$ can be written as [23, 24]

$$n_e^2(\theta) = \left(\frac{\cos^2\theta}{n_{e0}^2} + \frac{\sin^2\theta}{n_o^2}\right)^{-1} \quad (2)$$

where $n_{e0}$ is $n_e(\theta)$ at $\theta = 0$ and $n_o$ is the refraction index of ordinary light. Then, the phase retardation of light for parallel aligned LC cells is [23, 24]

$$\Delta\phi = \frac{2kdn_{e0}n_o}{\sqrt{n_{e0}^2 \cos^2\left[2\tan^{-1}\exp\left(\frac{V_C-V}{V_0}\right)\right]+n_o^2 \sin^2\left[2\tan^{-1}\exp\left(\frac{V_C-V}{V_0}\right)\right]}} \quad (3)$$

where $k = 2\pi/\lambda$ is the wavenumber of incident beam. The relationship between the applied voltage and the displayed grey level ($g$) is expressed as

$$V = V_L + (V_H - V_L)\frac{g}{g_{max}} \quad (4)$$

where $g_{max}$ is the grey-level number of the SLM. According to Jones matrix analysis, the reflection light intensity ($I$) during phase modulation of SLM is related to $\Delta\phi$ as

$$I = (\sin\psi + \cos\psi\cos\Delta\phi)^2 + (\sin\Delta\phi\cos\psi)^2 \quad (5)$$

where $\psi$ is the incident angle of linearly polarized light. If $\psi = 0$, the LC SLM only works in phase modulation mode with $I = 1$. If $0 < \psi < \pi/2$, LC SLM works in both phase modulation and intensity modulation. If $\psi = \pi/2$, there is neither phase modulation nor intensity modulation. Ideally, the phase-only SLM should be employed at $\psi = 0$. However, $\psi$ is slightly larger than zero in practical application. Thus, according to Eq. (3-5), $V_H$ and $V_L$ determines $I$ and $\Delta\phi$, which in turn influence the intensities of the modulation light ($I_m$) and the zeroth-order light ($I_z$). The modulation efficiency ($\delta$) can be evaluated as

$$\delta = \frac{\iint_{\Omega_1} I_m(V_H,V_L)d\Omega_1}{\iint_\Omega I(V_H,V_L)d\Omega} \times 100\% \quad (6)$$

with

$$\iint_\Omega I(V_H,V_L)d\Omega = \iint_{\Omega_1} I_m(V_H,V_L)d\Omega_1 + \iint_{\Omega_2} I_z(V_H,V_L)d\Omega_2 + P_{loss} \quad (7)$$

where $I_m(V_H,V_L)$ is the modulated light intensity at different $V_H$ and $V_L$, $I_z(V_H,V_L)$ is the zeroth-order light intensity. $\Omega_1$, $\Omega_2$ and $\Omega$ are the integral region of $I_m$, $I_z$ and $I$ respectively. $P_{loss}$ denotes the light power loss. The portion of the zeroth light is accordingly

$$\eta = \frac{\iint_{\Omega_2} I_z(V_H,V_L)d\Omega_2}{\iint_\Omega I(V_H,V_L)d\Omega} \times 100\% \quad (8)$$

In practical applications, the larger the $\delta$, the higher the light intensity of the modulation beam, and accordingly, the lower portion of the zeroth light if $P_{loss}$ is approximately constant.

The influence of $V_L$ on phase retardation $\Delta\phi$ varying with $V$ is shown in Fig. 1(a), where $n_{e0} = 1.74$, $n_o = 1.51$, $V_H = 4.19$ V and $d = 2.7865$ μm. When $V_L$ is increased, although the range of $\Delta\phi$ is reduced, the linearity of the modulation curve is increased, with improved modulation accuracy. This is important to avoid the zeroth-order light due to modulation deviation. In contrast, a larger $V_H$ is accompanied by a larger range of $\Delta\phi$ (Fig. 1(b)).

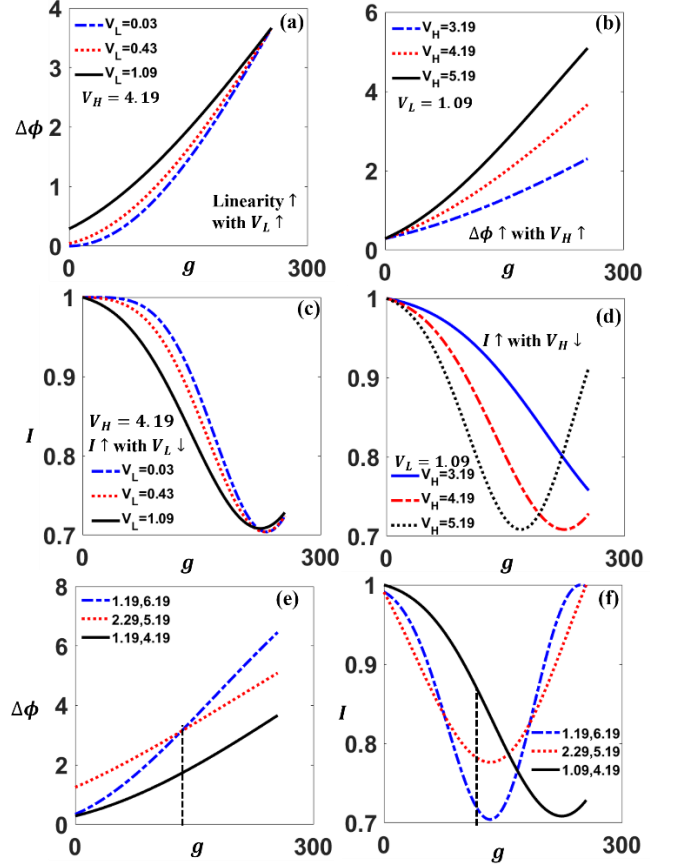

**Fig. 1.** Change of $\Delta\phi$ and $I$ under different parameters, where $d = 1.3891$ μm, $n_e = 1.74$, $n_o = 1.51$. (a) Relationship between $g$ and $\Delta\phi$ under different $V_L$. (b) Relationship between $g$ and $\Delta\phi$ under different $V_H$. (c, d) Influence of $V_L$ and $V_H$ on $I$ when $\psi = 5°$. (e, f) $\Delta\phi$ and $I$ under the optimal $V_L$ and $V_H$.

The phase retardation has a strong influence on the reflection light intensity, as inferred by Eq. (5), if the LC works at an inevitable angle $\psi \neq 0$. Here, we take $\psi = 5°$ as an example. From Fig. 1(c, d), it can be seen that the reflection light intensity $I$ clearly decreases with both $V_L$ and $V_H$.

In the application of SLM, researchers always prefer a higher modulation accuracy and linearity, in the meanwhile, with a higher $\delta$ and larger $I_m$. From Fig. 1(a-d), on one hand, increasing $V_L$ can promote the modulation accuracy and accordingly $\delta$. On the other hand, it decreases $\delta$ due to smaller $I$ (i.e. $I_m$). While decreasing $V_H$ leads to a smaller range of $\Delta\phi$, but a larger $I_m$. By optimizing $V_L$ and $V_H$, the compromise between the modulation efficiency and reflection intensity can be accommodated with both high modulation accuracy and $\delta$. From Fig. 1(e, f), it can be seen under an optimal $V_L = 1.09$ and $V_H = 4.19$, a better modulation accuracy has been realized. The reflection light intensity under the optimal condition also exhibits the highest value, in a widely used phase modulation range of $g = 0 \sim 127$. Thus, by properly selecting $V_L$ and $V_H$, $I_m$ could be improved for better $\delta$, with sufficiently good modulation accuracy.

To evaluate the effect of the method, a relative modulation efficiency $\delta_r$ is defined as following

$$\delta_r = \frac{\delta(V_H, V_L) - \delta_0}{\delta_0} \times 100\% \quad (9)$$

where $\delta_0$ is the modulation efficiency of the default $V_L$ and $V_H$. $\delta_r > 0$ indicates an enhanced modulation efficiency.

In Fig. 2, we show the numerical simulation of the light intensity distribution after modulation with different $V_H$ and $V_L$. The simulation is carried out by Richards-Wolf vectorial diffractive theory [25] for an aplanatic high-NA objective lens obeying the sine condition (see supplementary materials). Two typical vectorial polarization beams, i.e., perfect vortex beam (Fig. 2(a-f)) and Airy beam (Fig. 2(a1-f1)), are investigated. Fig. 2(a) shows the modulation result when $V_H$ and $V_L$ are default values. As $V_H$ is decreased, $\delta_r$ are all positive indicating improved $\delta$ (Fig. 2(b, c)). The improvement is relative limited ($\delta_r \leq 2\%$) indicating larger phase modulation capability through appropriately increasing $V_H$ has a minor effect on $\delta$.

In contrast, $\delta_r$ can be rapidly increased with $V_L$, as shown in Fig. 2(d-f). The maximum $\delta_r$ is 9.5% and the corresponding $\delta = 94.6\%$. The improvement of the modulation effect, as can be seen from the simulation, is remarkable according to the clear inhibition of diffraction fringes. Similar results can also be found in the modulation of Airy beams, where the maximum $\delta_r$ and $\delta$ are 8.2% and 93.2% respectively.

The observations above are further experimentally validated in an inverted fluorescence microscope (Fig. 3(g)). Here, we used a reflective phase-only LC SLM (LETO, HOLOEYE Photonics AG, Germany, PLUTO-VIS-056, 420 nm~650 nm) as an example to show the influence of $V_H$ and $V_L$ on light modulation. The phase maps of the modulation beams are designed by a phase modulation function (see supplementary materials), accompanied by a blazed grating.

Perfect vortex beam and Airy beam are generated under different $V_H$ and $V_L$. Fig. 3(a-f) show the light intensity distributions of

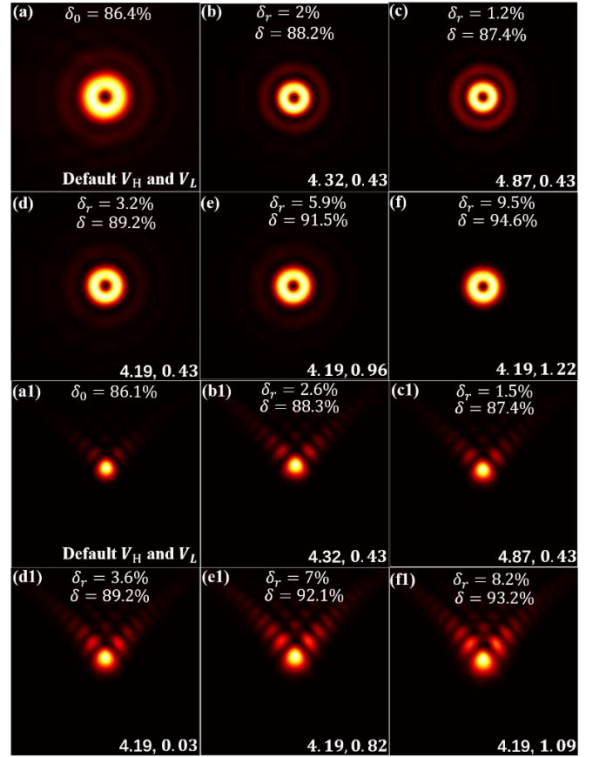

**Fig. 2.** Influence of $V_H$ and $V_L$ on the modulation of vectorial beams by numerical simulation. The NA of the objective lens is 1.4. The wavelength is 639 nm. Here, a standard axicon phase and a cubic phase are used to generate a perfect vortex beam and Airy beam. $V_H$ and $V_L$ are provided in the bottom right of the figures in sequence. (a-f) Light intensity of perfect vortex beams under different $V_H$ and $V_L$. (a) $V_H = 6.19$ and $V_L = 0.03$ are the default values from the manufacturer. (a1-f1) Light intensity of Airy beams under different $V_H$ and $V_L$. (a1) $V_H = 6.19$ and $V_L = 0.03$ are the default values from the manufacturer.

perfect vortex beams. With fixed $V_H$ but increased $V_L$, the zeroth-order light intensity gradually decreases from $\eta = 14.4\%$ to 1.3%, inhibited up to 91.3%. Accompanied, the modulated perfect vortex beam becomes brighter with an increasing $\delta$ from 77.5% to 92.6%. Relative to the default values of $V_H$ and $V_L$, the relative modulation efficiency $\delta_r = 19.5\%$. Similar results can also be observed in Fig. 3(a1-f1), where an Airy beam is generated. The zeroth-order light can be inhibited up to 83.6% with $\eta$ decreased from 11.3% to 1.9%. The zeroth-order light is weakening with the increasing $V_L$, accompanied by $\delta$ increasing from 74.1% to 92.1%. In contrast to $V_L$, decreasing $V_H$ can also improve $\delta$ with a minor effect (Fig. 3(b-d) and (b1-

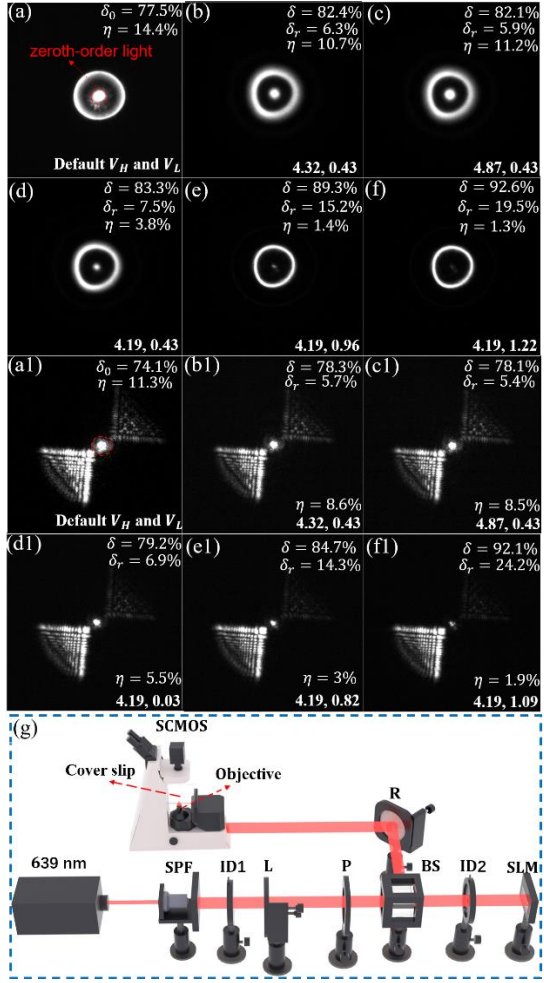

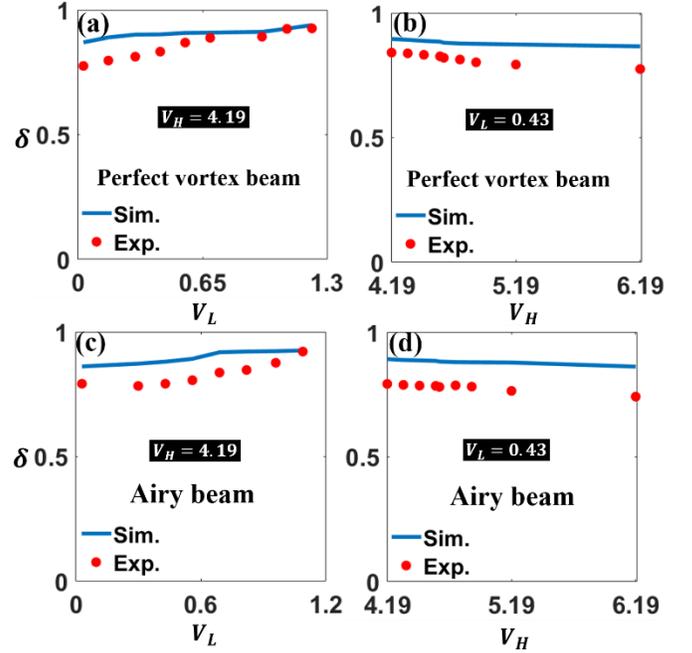

Fig. 3. Experimental setup and modulated focal spots. $V_H$ and $V_L$ are provided in the bottom right of the figures in sequence. (a-f) Fluorescent images of perfect vortex beam at different $V_H$ and $V_L$. Relative to (a), where $V_H = 6.19$ and $V_L = 0.03$ are the default values, an inhibition of the zeroth-order light is apparently observed. (a1-f1) Fluorescent images of Airy beam at different $V_H$ and $V_L$. In (a1), the same default $V_H$ and $V_L$ are applied. (g) Schematic of the optical system. A 639 nm continuous-wave laser is applied as light source. SPF is spatial pinhole filters; ID1 and ID2 are iris diaphragm; L is a collimation lens; P is a polarizer; BS is a beam splitter; SLM is a liquid-crystal-on-silicon spatial light modulator; WP is a quarter-wave plate; R is reflective mirrors. Here, fluorescent image is captured with high-NA objective lens (Nikon PL APO 63X NA 1.4 oil-immersion).

d1)) The experimental results are qualitatively consistent with that in numerical simulations, and clearly support that, by optimizing $V_L$ and $V_H$, $\delta$ can be improved, accompanied by the inhibition of the undesired zeroth-order light.

Fig. 4. Numerical simulation and experimental results of $\delta$ when modulating different beams to reveal the influence of $V_L$ and $V_H$ of SLM. (a, c) Influence of $V_L$ on $\delta$ for perfect vortex beam and Airy beam respectively. Here, $V_H = 4.19$ V. (b, d) Influence of $V_H$ on $\delta$ for perfect vortex beam and Airy beam respectively. Here, $V_L = 0.43$ V.

Fig. 4 shows how $\delta$ is influenced by $V_H$ and $V_L$ in both numerical simulations and experiments. It can be seen, with fixed $V_H = 4.19$ V but increased $V_L$, the $\delta$ gradually increases for both the perfect vortex beam and Airy beam as plotted in Fig. 4(a) and (c). The experimental and numerical results show acceptable consistency, e.g., when $V_L$ is 1.22 V, $\delta$ of the perfect vortex beam in numerical simulation and experiment are 94.6% and 92.6% respectively. While in Fig. 4(b, d), with fixed $V_L = 0.43$ V but increased $V_H$, $\delta$ slowly decreases for both the perfect vortex beam and Airy beam. Moreover, the optimal $V_H$ and $V_L$ to reach the highest $\delta$ rely on the $g$ range of the designed beam, and the accompanied range of $V$. If the range of $g$ to modulate the beam is located in the high $I$ region of Fig. 1(b, d), $\delta$ could be significantly increased.

To this end, we have shown a simple but effective approach for eliminating zeroth-order light caused by the pixelation effect of SLM in light modulation. With the appropriate selection of the $V_L$ and $V_H$ voltages, the zeroth-order light can be significantly inhibited and the modulation efficiency of SLM is accordingly improved. In experiments, the zeroth-order light can be inhibited up to 91.3%, accompanied by an improvement of the modulation efficiency from 76% to 92.6%. We hope the current method will have a promising influence on the development of SLM-based techniques for high-performance laser fabrication, optical imaging and tweezers etc.

**Funding.** This investigation is supported by National Natural Science Foundation of China (Grant No. 51927804, 61775181, 61378083).

**Disclosures.** The authors declare no conflicts of interest

See Supplement 1 for supporting content.